 \documentclass[final,5p,times,twocolumn,authoryear]{elsarticle}

\usepackage{enumerate}
\usepackage{comment}
\usepackage{amsmath}
\usepackage{amssymb}
\usepackage{amsfonts}
\usepackage{physics}
\usepackage[dvipsnames]{xcolor}
\usepackage{feynmp-auto}
\usepackage{epsfig}
\usepackage{slashed}
\allowdisplaybreaks
\usepackage{blindtext}
\usepackage{hyperref}
\usepackage{setspace}
\usepackage{tikz}
\usepackage{lipsum}
\usepackage{ragged2e}
\usepackage{xcolor}
\usepackage{soul}
\usepackage[c]{esvect}
\usepackage{harpoon}
\usepackage{subcaption}
\usetikzlibrary{angles,quotes}
\usepackage[english]{babel}
\usepackage[autostyle]{csquotes}
\usepackage{nicematrix}
\usepackage{blkarray}

\usepackage{amsthm}
\newcommand{\bc}{\begin{center}}
\newcommand{\ec}{\end{center}}

\renewcommand{\bf}[1]{\mathbf{#1}}

\newcommand{\ra}{\rangle}
\newcommand{\la}{\langle}
\newcommand{\nn}{\nonumber}

\begin{document}

\begin{frontmatter}



\title{Unitarity constrains the quantum information metrics for particle interactions}

%\author[label1,label2]{}

\author[First]{Shanmuka Shivashankara}
\affiliation[First]{organization={Providence College},            addressline={1 Cunningham Square}, city={Providence}, postcode={02918}, state={RI},country={USA}}

\author{Hobbes Sprague}
%\affiliation[Second]{ city={Providence}, state={RI},country={USA}}

\begin{abstract}
Unitarity provides mathematical and physical constraints on quantum information systems.  e.g., in entanglement swapping, unitarity requires the same von Neumann entanglement entropy generation for either a particle interaction or an act of measurement. 
  For the first time, the language of non-relativistic quantum mechanics is presented to derive the density matrix for hard scattering.  We show that unitarity allows for finding the latter's cross section without using the scattering amplitude or the Lippmann-Schwinger equation plus Green's function.  We also show the language of relativistic quantum mechanics can be used to derive the momentum entropy or Sackur-Tetrode equation for the inelastic scattering of an electron from a proton.  The latter entropy derives from a Shannon entropy and an additional entropy that evokes the uncertainty principle.  This article's presentation allows particle physicists to readily begin calculating quantum information metrics such as correlations and mutual information for any particle interaction.         
\end{abstract}




\end{frontmatter}

%\tableofcontents



\section{Introduction}\label{intro}

Recently, entanglement was found in a top-antitop quark pair by the ATLAS Collaboration at the CERN-LHC \cite{atlas}.  Since the top-antitop quark pair decays before hadronization, the daughter leptons convey entanglement information about their parent quarks.  As particle physicists become more interested in adding quantum information metrics such as entanglement entropy to their toolkit, having a brief primer on the importance of unitarity in calculating density matrices will be helpful.  The density matrix is the essential ingredient for calculating entanglement entropies, correlations, expectation values, mutual information (separability) of particles' degrees of freedom such as spin and momenta. 

In section \ref{unitarity}, unitarity is shown to provide mathematical and physical constraints on particle interactions such as a decay and scattering process.  A  primer on the constraints of unitarity is essential since the published literature has been omitting unitarity.  e.g., in \cite{Blasone}, one particle in an entangled pair partakes in Bhabha scattering.  The other particle or witness in the entangled pair does not participate in the interaction.  Since the authors do not maintain unitarity, the witness particle's reduced density matrix changes if its entangled partner undergoes parity violating interactions.  Unitarity forbids this outcome. The violation of unitarity stems from allowing the normalization of the density matrix to change after an interaction.  See items ($\ref{i}$) and ($\ref{v}$) in section \ref{unitarity}.         

In the case of entanglement swapping, there exists initial entanglement within  two separate pairs of particles, but no entanglement between the pairs.  If an interaction occurs between one particle from each pair, the remaining two particles can acquire a nonzero entanglement entropy.  These remaining two particles would undergo the same von Neumann entanglement entropy generation if the other two particles had been measured instead of being allowed to interact directly with one another.  Without unitarity, this would not necessarily be true.  See item ($\ref{iv}$) in section \ref{unitarity}.       

In section \ref{hardscatt}, the language of non-relativistic quantum mechanics is used to present the algorithm for calculating the density matrix for hard scattering.  Without the scattering amplitude or using the Lippmann-Schwinger equation plus Green's function, unitarity and density matrices are sufficient for deriving the scattering cross section. In section \ref{inclusive}, for the inelastic process $e^-p\rightarrow e^-\sum\limits_i X_i$, we find the electron's Sackur-Tetrode equation or momentum entanglement entropy has two contributions.  One contribution is due to the Shannon entropy for inelastic scattering $to\ occur$ or $not\ to\ occur$ while the other contribution evokes the uncertainty principle.

\section{Unitarity  constrains the final density matrix and von Neumann entanglement entropy}\label{unitarity}
Let $|i\ra$ represent the initial state. After an interaction, e.g. a decay or scattering process, the final state is $|f \ra = S|i\ra$,  where $S$ is the unitary $S$-matrix.  The initial and final density matrices are $\rho^i = |i\ra \la i|$ and $\rho^f = |f \ra \la f |$, respectively. Preserving unitarity imposes constraints on the density matrix as follows in items $(\ref{i})$ through $(\ref{vii})$ below.
\begin{enumerate}[$(i)$]

\item \label{i} \textit{Unitarity preserves the normalization.}

Taking the trace of the final density matrix,
\begin{align*}
Tr(\rho^f) &=  Tr(S\rho^i S^\dagger)=Tr(S^\dagger S \rho^i)\\ 
&= Tr( \rho^i).
\end{align*}
Hence, unitarity does not allow an interaction, e.g. a scattering or decay process, to $leak$ probability from the system.  

In \cite{seki}, they consider an elastic scattering process.  By writing their density matrix's normalization in terms of final states and doing a perturbative expansion, unitarity is lost.  In their eqn. (2.17), their coefficient in front of the momentum matrix position $|p_1\ra\la p_1|$ should be the probability for no scattering since the initial state survives. See eqn. (\ref{genrho})  in item $(\ref{iii})$ below and its interpretation.  More recently, in \cite{araujo} and \cite{Blasone}, unitarity is dropped allowing the normalization of a density matrix to change after a particle interaction.

\item\label{ii} \textit{Unitarity preserves the purity of a system.}

The purity is given by the trace of the square of the density matrix.  After an interaction, the final density matrix has the form $\rho^f = S\rho^i S^\dagger$ with a purity
\begin{align*}
Tr(\rho^{f\ 2}) &= Tr( S\rho^i S^\dagger S\rho^i S^\dagger) = Tr( S\rho^{i\ 2} S^\dagger) \\
&=Tr(\rho^{i\ 2}).
\end{align*}
Therefore, unitarity preserves the purity after an interaction.  If $Tr(\rho^2) = Tr(\rho)$, then the matrix is called pure.  Otherwise, it is impure.  For a pure system, the von Neumann entanglement entropy, $S^{EE}=-Tr(\rho\log\rho)$, is zero.  This means the known information about the system is maximal.  With unitarity maintained, the system would still have maximal information after an interaction.

\item\label{iii} \textit{Unitarity implies the general regularized form of a final density matrix \citep{shiva3}.}

  Let $|i\rangle$ and $S$ be the initial state and unitary $S$ matrix, respectively.  After an interaction, the final state is $|f \ra= \dfrac{1}{N}\ S|i\rangle = \dfrac{1}{N}\ (1+ i\mathcal{T}) |i\rangle$ with a normalization $N=\sqrt{\langle i|i \rangle}$ and $\mathcal{T}$ being the transition operator.
Expanding out the final density matrix, $\rho^f = |f\rangle \langle f| = \dfrac{1}{N^2}\ S | i \rangle \langle i | S^\dagger$, gives $\rho^f= \dfrac{1}{N^2}\ \Big(|i\rangle\langle i|+ (i \mathcal{T})|i\rangle\langle i|+H.c. + (i\mathcal{T})|i\rangle\langle i|(-i\mathcal{T}^\dagger)\Big)$.
Keeping the momenta and polarizations suppressed, insert the identity operator of final states ($\sum\limits_{\alpha} \dfrac{1}{\langle \alpha|\alpha \rangle}\ |\alpha\rangle \langle \alpha|$) next to all the transition operators, $\mathcal{T}$, and group terms, giving
\begin{align*}
\rho^f= &\dfrac{1}{N^2} \Big(|i\rangle \langle i|+ \langle i| (i \mathcal{T})|i\rangle\ \frac{|i \rangle \langle i|}{N^2}+H.c. + \sum_{\alpha \neq i} \langle \alpha | (i \mathcal{T})|i\rangle\ \dfrac{|\alpha \rangle \langle i|}{\langle \alpha|\alpha\rangle}\\
&+H.c.  +  \sum_{\alpha,\alpha'} \dfrac{1}{\langle \alpha|\alpha \rangle}\ \langle \alpha |(i\mathcal{T})|i\rangle\langle i|(-i\mathcal{T}^\dagger)|\alpha'\rangle \ \dfrac{|\alpha \rangle \langle \alpha'|}{\langle \alpha'|\alpha' \rangle} \Big)
\end{align*}
The second and third terms above can be cast into the imaginary part of the Feynman amplitude, $\mathcal{M}(\cdot)$.
\begin{align*}
\langle i | (iT) |i \rangle\ +\ H.c.\ =
-2VT\ Im\ \mathcal{M}( i \rightarrow\ i)
\end{align*}
The unregularized volume and time above are $V=(2\pi)^3\delta^3(0)$ and $T=2\pi\delta(0)$, respectively.  By the optical theorem, $Im \ \mathcal{M}(i\rightarrow i)$ above is proportional to the total decay width for  a decay process or the total scattering cross section for a scattering process. The general form of the density matrix becomes
\begin{align}\label{genrho}
\rho^f= &\ \Big(1 - \dfrac{2Im\ \mathcal{M}(i\rightarrow i) }{\la i | i \ra/VT}\Big) \ \dfrac{|i\rangle \langle i|}{\la i | i \ra}\nn\\
&+ \frac{1}{\la i | i \ra}\sum_{\alpha \neq i} \langle \alpha | (i \mathcal{T})|i\rangle\ \dfrac{|\alpha \rangle \langle i|}{\langle \alpha|\alpha \rangle}+H.c.\nn\\
&+  \dfrac{1}{\la i | i \ra}\sum\limits_{\alpha,\alpha'} \dfrac{1}{\langle \alpha|\alpha \rangle}\ \langle \alpha |(i\mathcal{T})|i\rangle\langle i|(-i\mathcal{T}^\dagger)|\alpha'\rangle \ \dfrac{|\alpha \rangle \langle \alpha'|}{\langle \alpha'|\alpha' \rangle}.
\end{align}
Notice the type of interaction, e.g. a decay or scattering process, was not specified in deriving $\rho^f$ above.\\
To interpret $\rho^f$, assume the complete trace of $\rho^f$ is a sum of probabilities. Then, the matrix elements associated with $\dfrac{|\alpha\rangle\langle \alpha' |}{\langle \alpha'| \alpha' \rangle}$ in the above final density matrix gives probabilities of final states occurring with specific degrees of freedom such as momenta and polarizations.  Therefore, the coefficient in front of $\dfrac{| i \rangle\langle i |}{\la i | i \ra}$ in the final density matrix must be the probability of no interaction, e.g. no scattering or no decay of the initial state.  Setting this probability to zero provides the final density matrix assuming an interaction occurs.  Setting it to zero is the $S$-matrix perspective, $i.e.$ in the remote future.  Therefore the $S$-matrix regularization is
\begin{align}\label{regansatz}  \dfrac{\la i|i \ra}{VT} = 2Im\ \mathcal{M}(i\rightarrow i).    
\end{align}
After writing eqn. (\ref{genrho}) in terms of the Feynman amplitudes, the above ratio, $\dfrac{\la i | i \ra}{VT}$, appears in all matrix elements as a common denominator.

The above regularization can be relaxed for a finite time such as a decay process.  Consider the decay of a polarized muon of mass $m_\mu$.  The initial state's normalization is $\la i| i\ra = 2m_\mu V$.  From eqn. (\ref{genrho}) and the optical theorem, the probability for no decay at time $t$ is $1 - \dfrac{2Im\ \mathcal{M}(i\rightarrow i) }{\la i | i \ra/VT} = 1 - \dfrac{Im\ \mathcal{M}(i\rightarrow i)}{m_\mu} T= 1 - \Gamma T$, where $\Gamma$ is the total decay width of the muon.  This probability of survival should be $e^{-\Gamma t}$. To prove the latter, write the unregularized time, $T\equiv2\pi\delta(0) = \lim \limits_{E_f\rightarrow E_i} \int_{-\infty}^{\infty} e^{i(E_f-E_i)t} dt$, with an imaginary energy term, $i.e.$ $E_f - E_i \rightarrow E_f - E_i + i\Gamma$. After integrating, the regularized time is $T=\dfrac{1-e^{-\Gamma t}}{\Gamma}$ and $ 1-\Gamma T = e^{-\Gamma t}$. As $t\rightarrow \infty$, this latter probability for no decay goes to zero, which is the $S$-matrix perspective.  

The latter implies a reduced final neutrino helicity ($\lambda$) density matrix consisting of the probability for a muonic decay and no decay \cite{shiva2}.  Furthermore, the von Neumann entanglement entropy, $S^{EE}_\lambda$, of the neutrino helicity is
\begin{align*}
S^{EE}_\lambda =& -Tr(\rho^f_\lambda \log \rho^f_\lambda)\\
=& -e^{-\Gamma t}\log e^{-\Gamma t} - (1-e^{-\Gamma t})\log (1-e^{-\Gamma t}).
\end{align*}

%\begin{comment}
\begin{figure}[htp]
	\centering 
	\includegraphics[width=0.45\textwidth, angle=0]{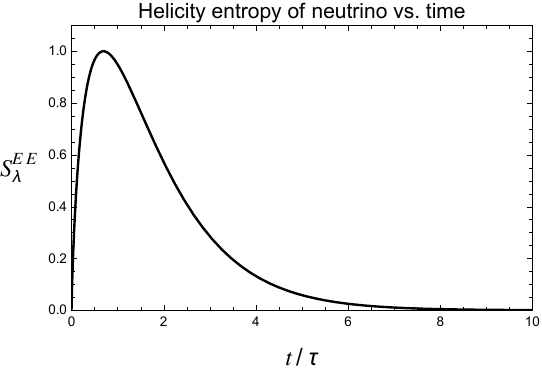}	
	\caption{Entropy $vs.$ time, $t$, in units of the muon's lifetime, $\tau$.  The von Neumann entanglement entropy of the neutrino's helicity, $S^{EE}_\lambda$, rises and falls with time since the birth of the parent-muon at $t=0$.  As $t\rightarrow \infty$, the entropy goes to zero since the polarized muon decays and the neutrino can only have one helicity. }\label{SEE}
\end{figure}
%\end{comment}
Assume the base of the above log is two.  The maximum entropy occurs at about two thirds the muon's lifetime when the probability for decay equals the probability for survival, $i.e.$ $S^{EE}_\lambda=\log 2 = 1$.  Figure (\ref{SEE}) graphs the above entropy with respect to time where a unit time is the lifetime, $\tau$, of the muon.  Notice as $t\rightarrow \infty$, $S^{EE}_\lambda \rightarrow 0$.  This means the muon definitely decays in the remote future and that the neutrino has a definite helicity.    However, the total entropy must be zero at all times because of unitarity and the initial state being pure.  \cite{shiva2} found the kinematic and angular $entropy$ distributions of the daughter electron to be similar to the corresponding decay rate distributions in the remote future.

\item\label{iv} \textit{Consider two pairs of particles.  Suppose there exists entanglement within each pair but not between the pairs.  Unitarity implies that the measurement of one particle from each pair or a direct interaction between the same two particles generates the same von Neumann entanglement entropy between the remaining two particles.  This is related to entanglement swapping.}

proof:
Let the initial state be\[
| i \ra = (a_1\ | R, \downarrow \ra + b_1\ | L, \uparrow \ra) \otimes (a_2\ | \downarrow, R \ra + b_2\ | \uparrow, L \ra)
\]
where momenta are suppressed and 
\[
|a_i|^2 + |b_i|^2 = 1, \quad i = 1,2.
\]
Both $R$ and $\uparrow$ ($L$ and $\downarrow$) refer to right (left)-handed helicities. For example, the term $a_1a_2|R, \downarrow \ra \otimes | \downarrow, R \ra$ refers to particles 1 and 4 (2 and 3) being right (left)-handed.

Suppose particles 2 and 3 interact.  The density matrix of particles 1 and 4 can be obtained as follows.  The final state is $| 
f \ra = S | i \ra$ with a final density matrix
\begin{align*}
p^f =&\ |f \ra \la f| = S | i \ra \la i | S^{\dagger}\\
=&\  S |\downarrow, \downarrow \ra \la \downarrow, \downarrow | S^{\dagger}\ |a_1a_2|^2 |R, R \ra \la R, R|  + \\
&\  S |\downarrow, \uparrow \ra \la \downarrow, \uparrow | S^{\dagger}\ |a_1 b_2|^2 |R, L \ra \la R, L|  +\\
&\ S |\uparrow, \downarrow \ra \la \uparrow, \downarrow | S^{\dagger}\ |b_1 a_2|^2 |L, R \ra \la L, R|  +\\
&\ S |\uparrow, \uparrow \ra \la \uparrow, \uparrow | S^{\dagger}\ |b_1 b_2|^2 |L, L \ra \la L, L|
+ 12 \text{ coherence terms}.
\end{align*}
  Let $S \to IS$, where $I \equiv \sum\limits_\alpha |\alpha\ra \la \alpha|$ is the identity operator of final states for the interaction between particles 2 and 3. Then, trace over the final states of particles 2 and 3. For example, consider tracing the first term in $\rho^f$ above.
\begin{align*}
&Tr_\alpha \Big( \sum\limits_\alpha |\alpha \ra \la \alpha|\ S | \downarrow, \downarrow \ra \la \downarrow, \downarrow | S^{\dagger}\Big)\  |a_1 a_2|^2 |R, R \ra \la R, R|\\
&= \Big( \sum\limits_\alpha \la \alpha | S | \downarrow, \downarrow \ra \la \downarrow, \downarrow  | S^{\dagger} |\alpha\ra \Big)\  |a_1 a_2|^2 |R, R \ra \la R, R|\\
&= \Big( \la \downarrow, \downarrow | S^{\dagger} \sum\limits_\alpha |\alpha \ra \la \alpha |\ S | \downarrow, \downarrow \ra \Big)\ |a_1 a_2|^2 |R, R \ra \la R, R|&\\
&=|a_1 a_2|^2 |R, R \ra \la R, R|,
\end{align*}
where the factor in parentheses in the second-to-last line is one since
$I = \sum\limits_\alpha |\alpha \ra \la \alpha|,\ S^{\dagger} S = 1$, and $\la \downarrow, \downarrow | \downarrow, \downarrow \ra = 1$.

Therefore, tracing over the remaining 15 terms in \( \rho_f \) gives the final density matrix for particles 1 and 4.
\\
\[
\rho_{1,4}^f = 
\bordermatrix{
    & |R,R\ra & |R,L\ra & |L,R\ra & |L,L\ra \cr
    \la R,R| & |a_1 a_2|^2 & 0 & 0 & 0 \cr
    \la R,L| & 0 & |a_1 b_2|^2 & 0 & 0 \cr
    \la L,R| & 0 & 0 & |b_1 a_2|^2 & 0 \cr
    \la L,L| & 0 & 0 & 0 & |b_1 b_2|^2
}
\]

This is the same von Neumann entanglement entropy generation between particles 1 and 4 that is had by tracing (measuring) the initial density matrix,
$\rho^i = |i \ra \la i |$,
over particles 2 and 3. Also, the same mutual information (separability) between particles 1 and 4 is generated, namely zero.  The result holds even if the initial pairs are mixed states.  To obtain a nonzero mutual information, the final density matrix must be projected onto a particular final state such as a Bell state at which point the similitude of measurement and interaction no longer hold.   

\item\label{v} \textit{Suppose a pair of particles, say $x$ and $a$, are entangled.  The pair is not entangled with a third particle, $b$.  If particle $a$ interacts with $b$, $x$ is said to be a $spectator$ or witness since it does not participate in the interaction. Unitarity implies that the witness particle's reduced density matrix and von Neumann entropy are unchanged in spite of its entangled partner's interaction. However, $x$ and $b$ do become entangled after the interaction.}\\
The proof is similar to $(\ref{iv})$ above and given in \cite{shiva1}.  As an example of the witness particle's reduced density matrix possibly changing due to $not\ keeping$ unitarity, see \cite{Blasone} or \cite{araujo}.  In \cite{Blasone}, they study Bhabha scattering with a witness particle.  Although their witness' reduced density matrix is unchanged after the Bhabha interaction, this would not occur if parity-violating interactions were included. The latter violation would occur because their normalization of the density matrix changes after the interaction.  By dropping unitarity or the forward scattering amplitude, known physical results can be altered.

In \cite{shiva1}, they study Compton scattering with a witness photon, $x$, but do not regularize the divergent volume, $V$, and time, $T$.  The scattering particles $a$ and $b$ are a photon and an electron, respectively. Before the Compton interaction, only the witness photon, $x$, and scattering photon, $a$ are entangled.  Using eqn.(\ref{genrho}) and the optical theorem, the probability for no Compton scattering is $1 - \dfrac{2Im\ \mathcal{M}(i\rightarrow i) }{\la i | i \ra/VT} = 1 - \dfrac{2Im\ \mathcal{M}(i\rightarrow i)}{2E_a 2E_b\ \upsilon} \dfrac{1}{(V/\upsilon T)} = 1 -  \dfrac{\sigma}{(V/\upsilon T)}$,
where $\sigma \equiv \sum_f\sigma_{ab\rightarrow f}$ is the  total scattering cross section of particles $a,b$.  $\upsilon $ is the relative velocity between particles $a,b$. Setting this probability to zero gives the final density matrix assuming Compton scattering does occur as well as the $area$ regularization 
\begin{align}\label{comptonreg}
\dfrac{V}{\upsilon T} \equiv \sigma.
\end{align}
Hence, the interaction rate, $1/T$, divided by the luminosity, $\dfrac{\upsilon}{V}$, is the total scattering cross section.  The Thomson scattering cross section is the regularization in eqn. (\ref{comptonreg}) at low energy.  The regularization can be relaxed for a finite time, e.g. a resonance that decays into a final state.

Eqn. (\ref{comptonreg}) also occurs when using the language of nonrelativistic quantum mechanics. In the latter the identity operator of final states, the inner product of momentum states, the $S$-matrix are written differently relative to the language of quantum field theory.  Referring back to the final density matrix in eqn. (\ref{genrho}), trace over the initial particles and set the probability for no scattering to zero.  Then trace over all final particles, obtaining $1=\dfrac{\sum_{i} \sigma_i}{\dfrac{V}{\upsilon T}}$, confirming eqn. (\ref{comptonreg}).  The cross section does not change when switching from quantum field theory to non-relativistic quantum  mechanics.  

Eqn. (\ref{comptonreg}) implies a finite mutual information and correlation between particles' degrees of freedom.  The mutual information is defined as $S_1 + S_2 - S_{12}$, where $S_i$ and $S_{12}$  are the von Neumann entanglement entropies of particle  $i$ and the two-particle system, respectively.  Before Compton scattering, the correlation and mutual information between the electron's and witness photon's helicities are zero.    The mutual information being zero means the helicities have no entanglement, $i.e.$ the states are separable.  After the Compton interaction, this mutual information is nonzero since the witness photon became entangled with the electron.  However, the witness photon's reduced density matrix remains unaffected by the interaction.  

In general, the above regularization implies a relationship between the density matrix and differential scattering cross section.  By tracing eqn. (\ref{genrho}) over the initial particles and using eqn. (\ref{comptonreg}), one obtains
\begin{align}\label{rhosigma}
Tr(\rho^f)=\dfrac{1}{\sigma}\int d\Omega\dfrac{d\sigma}{d\Omega}=1.
\end{align}
Recently, \cite{kow} considered flavored scalar scattering.  Their final density matrix has an undetermined ratio of time to volume divergences that is shown to be bounded by considering wave packets in momentum space.          

\item\label{vi} \textit{Suppose the initial density matrix is normalized. The von Neumann entanglement entropies are non-negative, $S^{EE}\geq0,$ when keeping unitarity across an interaction.}

The von Neumann entropy may be written as $S^{EE}=-Tr(\rho^f \log \rho^f) = -\sum_i \lambda_i \log \lambda_i$, where $\lambda_i$ is an eigenvalue of $\rho^f$.   The proof comes from seeing these eigenvalues of the density matrix as probabilities.   For an arbitrary vector $|x\ra$, $\la x|\rho^f|x\ra = |\la x|S|i\ra|^2\geq 0$. Hence, $\rho^f$ is Hermitian and positive semi-definite, $i.e.$  has non-negative eigenvalues.  Since the final density matrix has a trace of one due to unitarity, all its eigenvalues, $\lambda_i$, are less than or equal to one.

\item\label{vii} \textit{Consider hard scattering of a spinless particle of mass $M$ from a stationary, impenetrable sphere.  The scattering amplitude is $f(\vb*k',\vb*k)$, where $\vb*k$ and $\vb*k'$ are the incident particle's  initial and final wave vectors, respectively.  Assume $|\vb*k'| = |\vb*k|$ and define the differential scattering cross section as $\dfrac{d\sigma}{d\Omega} = |f(\vb*k',\vb*k)|^2$.  The regularization eqn. ($\ref{comptonreg}$)  implies the following relationship between the transition matrix element, $\la \vb*k' |\mathcal{T} |\vb*k \ra$, and the scattering amplitude up to a phase.}
\begin{align}\label{matrixT}
\la \vb*k' |\mathcal{T} |\vb*k \ra=&\ \dfrac{\hbar^2}{(2\pi)^2 M}f(\vb*k',\vb*k)\ \delta(\dfrac{\hbar^2 k'^2}{2M}-\dfrac{\hbar^2 k^2}{2M})
\end{align}
proof:  See \ref{transition-deriv}
%the supplementary file%
for the proof. 
 Also, the unitary $S$-matrix operator, $S$=$1-2\pi i \mathcal{T}$, and eqn. (\ref{matrixT}) imply the optical theorem, $\sigma = \dfrac{4\pi}{k}Im(e^{i(\pi+\alpha)}f(\vb*k,\vb*k))$.  $e^{i\alpha}$ is the missing phase in eqn. \ref{matrixT} and found to be $-1$ at the end of subsection (\ref{orbital}).  Normally, eqn. (\ref{matrixT}) is derived by the Lippmann-Schwinger equation plus the Green's function from which our missing phase is negative one (see chapter 6 in \cite{sakurai}).  
\end{enumerate}
%%Had to end "\end{enumerate}" here, otherwise, spacing messed up.

%\item\label{viii}\textit{}

\section{Density matrices of hard scattering via partial waves} \label{hardscatt}

Hard-sphere scattering is evaluated in this section to illustrate the density matrix algorithm given in item $(\ref{iii})$ in section \ref{unitarity}.   The notation in this entire section follows \cite{sakurai} but our derivations are self-contained.  Typically, the Lippmann-Schwinger equation plus Green's function is used to find the scattering amplitude from which the scattering cross section is obtained.  We demonstrate that the language of density matrices can be more efficient.  As shown in subsection \ref{orbital}, the scattering amplitude is not needed to find the total scattering cross section or average orbital quantum number of a scattered particle.

\subsection{Density matrix and expectation value of momentum}\label{algorithm}

Suppose that an incident spinless particle of mass $M$ undergoes elastic scattering from a stationary, hard (impenetrable) sphere of radius $R$. The normalized initial density matrix of the incident particle is $\rho^i = \dfrac{|\vb*k\ra \la \vb*k |}{\delta^3(0)}$, where $\vb*k$ is a wave vector such that $\la \vb*k | \vb*k'\ra = \delta^3(\vb*k - \vb*k')$. After scattering, its final state is $|\vb*k'\ra = S|\vb*k\ra$, where $S=1-2\pi i \mathcal{T}$ is the unitary operator and $\mathcal{T}$ is the transition operator.\\
The final density density matrix is
\begin{align*}
\rho^f = S \rho^i S^\dagger = \rho^i - 2 \pi i \ \mathcal{T}\rho^i + \text{H.c.} + (2 \pi)^2\  \mathcal{T} \rho^i  \mathcal{T}^{\dagger}.
\end{align*}
Insert the completeness relation, $I = \int d^3 \vb*k' \, |\vb*k' \ra \la \vb*k' |$, next to each $\mathcal{T}$ above giving
\begin{align}\label{apprho}
\rho^f =& \rho^i - \dfrac{2 \pi i}{\delta^3(0)} \int d^3 \vb*k' \, \la \vb*k' | \mathcal{T} |\vb*k \ra \, |\vb*k' \ra \la \vb*k | + \text{H.c.}\nn \\
&+ \dfrac{(2 \pi)^2}{\delta^3(0)} \int d^3 \vb*k' \int d^3 \vb*k'' \, \la \vb*k' | \mathcal{T} |\vb*k \ra \, \la \vb*k | \mathcal{T}^{\dagger} |\vb*k'' \ra \, |\vb*k' \ra \la \vb*k'' |
\end{align}

The first three terms pertain no scattering while the last term pertains to final states after scattering. From eqn. (\ref{matrixT}) the forward transition matrix element up to a phase is 
\begin{align*}
\la \vb*k | \mathcal{T} |\vb*k \ra =&\ \dfrac{T}{(2 \pi)^3} \dfrac{\upsilon}{k} f(\vb*k, \vb*k),
\end{align*}
where $T = 2 \pi \hbar \, \delta(0)$ is the unregularized time and $\delta(0)$ has units of inverse energy.  $\upsilon = \dfrac{\hbar k}{M}$ is the particle's speed with respect to a stationary hard sphere.  Planck's constant $\hbar$ is included for doing calculations.

After tracing over the initial particle, the final density matrix becomes 
\begin{align}\label{rhoderive}
\rho^f =& 1 + 2 \pi i\ \dfrac{1}{V/(\upsilon T)} \dfrac{1}{k}\  (e^{i(\pi+\alpha)}f(\vb*k, \vb*k) - \text{H.c.})\nn \\
&+ \dfrac{(2 \pi)^2}{\delta^3(0)} \int d^3 \vb*k' \int d^3 \vb*k'' \left( \dfrac{\hbar^2}{(2 \pi)^2 M} \right)^2 f(\vb*k', \vb*k) \ \delta(E' - \dfrac{\hbar^2 k^2}{2M})\nn \\
& f(\vb*k'', \vb*k)^* \, \delta(E'' - \dfrac{\hbar^2 k^2}{2M}) \, |\vb*k' \ra \la \vb*k'' |.
\end{align}
$V$ above is the unregularized volume, $V = (2 \pi)^3 \delta^3(0)$. If the argument of the latter Dirac delta function has momentum instead of the wave number, the volume becomes $V = (2 \pi \hbar)^3 \delta^3(0)$.  Using the optical theorem, the first two terms in eqn. (\ref{rhoderive}) equal $1 - \dfrac{\sigma_T}{V/(\upsilon T)}$, where $\sigma_T=\int d\Omega' |f(\vb*k',\vb*k)|^2$ is the total scattering cross section.  The unregularized final density matrix of the scattered particle has the form
\begin{align}\label{compare}
\rho^f=& 1 - \dfrac{\sigma_T}{V/(\upsilon T)} + \dfrac{1}{V/(\upsilon T)} \int d^3 \vb*k' \int d^3 \vb*k'' \dfrac{\delta(|\mathbf{k'}| - |\mathbf{k}|)}{|\mathbf{k}|^2}\nn\\
& *\frac{\delta(E'' - \dfrac{\hbar^2 k^2}{2M})}{\delta(0)}\ f(\vb*k', \vb*k) f(\vb*k'', \vb*k)^* |\vb*k' \ra \la \vb*k'' |.
\end{align}
The above density matrix has the form of a direct sum of the probability for no scattering, $1 - \dfrac{\sigma_T}{V/(\upsilon T)}$, plus probabilities of final states of the scattered particle. Let \( {V/(\upsilon T)} \equiv \sigma_T \), thereby obtaining the final density matrix such that scattering occurs. Notice \( {V/(\upsilon T)} \) has the form of an interaction rate, \( \dfrac{1}{T} \), divided by the luminosity, $\dfrac{\upsilon}{V}$, which gives the total scattering cross section. Our final density matrix of the scattered particle's momentum becomes
\begin{align}\label{rhotild}
\rho^f = \left( \delta^3(0) \int d^3 \vb*k' \, \delta^3(0) \int d^3 \vb*k'' \right) 
\dfrac{\mathcal{A}(\vb*k',\vb*k'')}{\sigma_T} \, \dfrac{|\vb*k' \ra \la \vb*k'' |}{\delta^3(0)},
\end{align}
where
\begin{align*}
\mathcal{A}(\vb*k',\vb*k'') \equiv  \dfrac{\delta(|\vb*k'| - |\vb*k|)}{\delta^3(0)}\ \dfrac{\delta(E'' - \dfrac{\hbar^2 k^2}{2M})}{\delta(0)}\ \dfrac{f(\vb*k', \vb*k) f(\vb*k'', \vb*k)^*}{k^2}\nn.
\end{align*}
The operator in parentheses in eqn. (\ref{rhotild}) is the continuous analogue of a discrete sum, $\sum_{\vb*k',\vb*k}$. Notice $\dfrac{\mathcal{A}(\cdot)}{\sigma_T}$ is a ratio of areas and the matrix element in $\rho^f$ at the matrix position $\dfrac{|\vb*k' \ra \la \vb*k'' |}{\delta^3(0)}$.  $\delta^3(0)$ has units of length cubed.    Let's confirm eqn. (\ref{rhotild}) by calculating $\la |\vb*k'| \ra$, which should be $|\vb*k|$. 
\begin{align*}
\la |\vb*k'|\ra =& \text{Tr}(|\vb*k'| \ \rho^f)\\
=& \left( \delta^3(0) \int d^3\vb*k' \delta^3(0) \int d^3 \vb*k'' \right)|\vb*k'| \dfrac{\mathcal{A}(\vb*k',\vb*k'')}{\sigma_T} \dfrac{\la \vb*k'' | \vb*k' \ra}{\delta^3(0)}\\
=& \dfrac{|\vb*k|}{\sigma_T} \int d\Omega' |f(\vb*k',\vb*k)|^2\\
=& |\vb*k|
\end{align*}
Also, the final momentum entanglement entropy is zero as expected since $\rho^{f\ 2} = \rho^f$.  If the entropy was nonzero, the information-theoretic Sackur-Tetrode equation occurs.  Eqn. (\ref{rhotild}) is also confirmed by calculating the correct scattering amplitude, $f(\vb*k',\vb*k)$, at the end of the next subsection.

\subsection{Density matrix of the orbital quantum number}\label{orbital}
In this subsection, the language of density matrices and unitarity are shown to be sufficient to obtain the scattering cross section, the average orbital angular momentum quantum number, the scattering amplitude $f(\vb*k',\vb*k)$, and the missing phase in eqn. (\ref{matrixT}) in that order.  To obtain the cross section, we derive the orbital number's density matrix by applying the completeness relation, $\sum_{\ell,m}\int dE\ |E, \ell, m\ra \la E, \ell, m |$, onto the normalized initial density matrix $\rho^i=\dfrac{|\vb*k\ra \la \vb*k |}{\delta^3(0)}$. $E,\ell,$ and $m$ are the energy, the orbital quantum number, and the magnetic quantum number, respectively.  Since $\la E,\ell,m| E',\ell',m' \ra = \delta(E-E')\  \delta_{\ell,\ell'}\ \delta_{m,m'}$, the wave function is $\la E,\ell,m| \vb*k \ra = \dfrac{\hbar}{\sqrt{Mk}}\ \delta(E-\dfrac{(\hbar k)^2}{2M})\ Y_\ell^m(\hat{k})^*$.  After tracing over $E$ and $m$, the orbital number's initial density matrix becomes
\begin{align}\label{rhol}
\rho^i_\ell =  \dfrac{1}{V/(\upsilon T)}\dfrac{\pi}{k^2}\ \sum_{\ell,\ell^{'}} \sqrt{(2\ell+1)(2\ell'+1)}\ |\ell\ra \la \ell'|. 
\end{align}

Since the trace is still one, the $area$ regularization 
is 
\begin{align}\label{sigmai}
\sigma\equiv \dfrac{V}{\upsilon T} = \dfrac{\pi}{k^2} \  \sum_\ell (2\ell+1).
\end{align}\
The latter formula also appears in eqn. (2.17) in \cite{pesch2}, where it is derived by a mathematical identity of $\delta$-functions in spherical coordinates.  In their eqn. (2.27), $\sigma$ above refers to the total cross section, which includes elastic and inelastic modes.  We interpret the above initial regularization $\sigma$  as an area through which a flux of particles, $\dfrac{\upsilon}{V}$, travels at the rate $1/T$.  Do not confuse $T$ eqn. (\ref{sigmai}) with $T$ in eqn. (\ref{compare}).  The latter $T$ derives from the transition matrix element in eqn. (\ref{matrixT}).  

Notice eqn. (\ref{rhol}) consists of the pure state
$\sqrt{\dfrac{\pi}{\sigma\ k^2}}\ 
\sum_\ell \sqrt{(2\ell+1)}\ |\ell\ra$.
Next, apply the $S$ matrix to the latter equation.  Assuming the particle definitely scatters, the final state becomes 
\begin{align*}
|L\ra \equiv \sqrt{\dfrac{\pi}{\sigma\ k^2}}\ 
\sum_\ell \sqrt{(2\ell+1)}\ (S_\ell -1)|\ell\ra,
\end{align*}
where $S_\ell$ is a diagonal element of the $S$ matrix.  Since the orbital angular momentum for each partial wave ($\ell$) is conserved, $S_\ell$ must be unitary.  Hence, $S_\ell$ must be a phase, say $e^{2i\delta_\ell}$. The final density matrix of $\ell$ becomes
\begin{align}\label{purelf}
\rho^f_\ell \equiv &\ |L\ra \la L|\nn\\
=&\ \dfrac{4\pi}{\sigma\ k^2}\ 
\sum_{\ell,\ell'} \sqrt{(2\ell+1)(2\ell'+1)}\ e^{i(\delta_\ell-\delta_{\ell'})}\sin\delta_\ell\sin\delta_{\ell'}\ | \ell\ra \la \ell' |.
\end{align}
Due to the application of $(S_\ell -1)$ above, $1/T$ above in $\sigma= \dfrac{V}{\upsilon T}$ is reinterpreted as the interaction rate. This must be true since the third term in eqn. (\ref{compare}) is eqn. (\ref{purelf}) after changing the basis from $\vb*k$ to $\ell$. Furthermore,  the trace of $\rho_\ell^f$ must be one by unitarity, implying the regularization or total scattering cross section for hard scattering is
\begin{align}\label{hardcs}
\sigma \equiv \dfrac{V}{\upsilon T} = \dfrac{4\pi}{k^2}\ \sum_\ell (2\ell +1)\sin^2\delta_\ell.
\end{align}
Compare the above regularization with eqn. (\ref{sigmai})'s regularization before scattering.  Therefore, the norm, $\dfrac{V}{\upsilon T}$, or $unregularized$ area of the density matrix remains after an interaction, but its regularization changes.  If inelastic modes were added to eqn. (\ref{rhotild}) and (\ref{purelf}), $\sigma$ above would include elastic and inelastic contributions.  

To obtain the expected orbital number, $\la \ell \ra$, for hard-sphere scattering at high energy, use eqn. (\ref{purelf}).  Assume a cutoff for the large sum at $\ell_{max} = kR$, where $R$ is the impact parameter, and replace $\sin^2\delta_\ell$ with its average value of $1/2$. %since the neighboring orbital modes are orthogonal, i.e. $|\delta_\ell - \delta_{\ell+1}|=\pi/2$ (see \cite{sakurai}).
This gives 
\begin{align*}
\la \ell \ra =& Tr(\ell\ \rho^f_\ell)= \dfrac{4\pi}{\sigma k^2}\sum_\ell ^{\ell_{max}}\ \ell\ (2\ell+1)\sin^2\delta_\ell\\
\approx & \dfrac{2kR}{3}
\end{align*}
with a total cross section of $\sigma\approx 2\pi R^2.$
Also, the von Neumann entanglement entropy of the orbital quantum number is still zero after the scattering since the final state is pure.

Lastly, we derive the spherically symmetric scattering amplitude, $f(\vb*k',\vb*k)$.  By changing the basis in eqn. (\ref{rhotild}) to $|E,\ell, m=0\ra$, tracing over $E$ and $m$,  and equating this to $\rho^f_\ell$ in eqn. (\ref{purelf}), the matrix element  $(\ell, \ell')$ implies
\begin{align*}
\int d\Omega' f(\vb*{ k}', \vb*{ k}) Y_\ell^{m=0}(\vb*{\hat k'}) = \dfrac{2\sqrt{\pi(2\ell+1)}}{k} e^{i\delta_\ell}\sin\delta_\ell.     
\end{align*}
Expanding the above $f(\vb*k',\vb*k)$ in terms of the Legendre polynomials, the correct scattering amplitude is 
\begin{align}\label{f}
f(\vb*{ k}', \vb*{ k}) = \dfrac{1}{k}\sum\limits_{\ell=0}^\infty (2\ell+1) e^{i\delta_\ell} \sin\delta_\ell P_\ell(\cos\theta),
\end{align}
where $P_\ell(\cos\theta)$ is the Legendre polynomial and $\theta$ is the scattering angle of the incident particle.  Lastly, 
the missing phase in eqn. (\ref{matrixT}) is found to be $-1$ by using eqn. (\ref{hardcs}), eqn. (\ref{f}), and the optical theorem as given at the end of item $(\ref{vii})$ in section \ref{unitarity}. 
 Without unitarity or the optical theorem, which suggests the regularization, the calculations in this subsection would not be possible.

\begin{comment}
\begin{table}
\begin{tabular}{l c c c} 
 \hline
 Source & RA (J2000) & DEC (J2000) & $V_{\rm sys}$ \\ 
        & [h,m,s]    & [o,','']    & \kms          \\
 \hline
 NGC\,253 & 	00:47:33.120 & -25:17:17.59 & $235 \pm 1$ \\ 
 M\,82 & 09:55:52.725, & +69:40:45.78 & $269 \pm 2$ 	 \\ 
 \hline
\end{tabular}
\caption{Random table with galaxies coordinates and velocities, Number the tables consecutively in
accordance with their appearance in the text and place any table notes below the table body. Please avoid using vertical rules and shading in table cells.
}
\label{Table1}
\end{table}
\end{comment}

\section{Sackur-Tetrode equation for inelastic scattering}\label{inclusive}

In this section, field theory language is used to obtain the Sackur-Tetrode equation for an electron scattering from a stationary proton in the inelastic process $e^-p\rightarrow e^- \sum\limits_{i} X_i$.  The latter process has an  $inclusive$ cross section of $\sigma_{in}$.  The initial electron-proton state is an unpolarized (mixed) state of helicities: ($(R,R),(R,L),(L,R),(L,L)$).  Let $|\vb*l,s\ra$ represent a particle with momentum $\vb*l$ and helicity $s$.  Its inner product is $\la \vb*k,r|\vb*l,s\ra =2E_{\vb*l}\  (2\pi)^3\delta^3(\vb*l-\vb*k) \delta_{rs}$.  Calculating the electron's quantum information metrics  requires the final density matrix.  The procedure for obtaining the latter follows section \ref{hardscatt}'s algorithm with differences in notation.  The $S$-matrix is $S=1+i\mathcal{T}$. The multiple particle identity operator  is $I=\prod\limits_{i}\Big(\ \sum\limits_{s_i}\int \dfrac{d^{3}\vb*l_{i}}{2E_{\bf{\vb*l}_i} (2\pi)^3}|\vb*l_{i},s_{i}\rangle\langle \vb*l_{i},s_{i}|\ \Big)$ while the momentum is used in lieu of the wave number.

After scattering, the total final state electron's momentum density matrix becomes 
\begin{align}\label{total}
\rho=(1-\dfrac{\sigma_{in}}{\sigma_T})\oplus \dfrac{\sigma_{in}}{\sigma_T}\rho_k.
\end{align}
The first term, $1-\dfrac{\sigma_{in}}{\sigma_T}$, represents the probability that inclusive scattering will not occur.  The second term has the probability for inclusive scattering to occur ($\dfrac{\sigma_{in}}{\sigma_T}$) times the density matrix, $\rho_k$, where    
\begin{align}\label{erho}
\rho_k=& \dfrac{V}{(2\pi\hbar)^3}\int d^3\vb*k\  \dfrac{\upsilon(\vb*k)}{V}\ \dfrac{|\vb*k\ra \la \vb*k|}{2E_{\vb*k}V}\\
\text{and}&\nn\\
\upsilon(\vb*k)=&\dfrac{(2\pi \hbar)^3}{\sigma_{in}}\dfrac{d\sigma_{in}}{d^3\vb*k}
%=\dfrac{1}{\sigma_{in}}\dfrac{\pi \hbar^5 c}{2E_{\vb*p} E_{\vb*k}} \dfrac{e^4 L^{\mu\nu}W_{\mu\nu}}{q^4}\nn.
\end{align} 
$\hbar$ has been inserted above to confirm dimensions.  The density matrix element $\upsilon(\vb*k)/V$ above is a ratio of volumes in the matrix position $\dfrac{|\vb*k\ra \la \vb*k|}{2E_{\vb*k}V}$.  The formula for $\dfrac{d\sigma_{in}}{d^3\vb*k}$ is given in eqn. (8.33) in chapter 8 in \cite{griffiths} assuming Bjorken scaling.  The correctness of $\rho_k$ and $\upsilon(\vb*k)$ above lies in $Tr(\rho_k)=1$.   From equations (\ref{total}) and (\ref{erho}), the electron's momentum entanglement entropy or Sackur-Tetrode equation is 
\begin{align}\label{see}
S^{EE}=&-Tr(\rho \log \rho)\nn\\
=&-(1-\dfrac{\sigma_{in}}{\sigma_T})\log(1-\dfrac{\sigma_{in}}{\sigma_T})-\dfrac{\sigma_{in}}{\sigma_T}\log(\dfrac{\sigma_{in}}{\sigma_T})
+\dfrac{\sigma_{in}}{\sigma_T}*S_k,
\end{align}
where
\begin{align}\label{sk}
S_k\equiv&-\log(2\pi\hbar)^3 + \log V - \int d^3\vb*k \dfrac{1}{\sigma_{in}}\dfrac{d\sigma_{in}}{d^3\vb*k}\log(\dfrac{1}{\sigma_{in}}\dfrac{d\sigma_{in}}{d^3\vb*k}).   
\end{align}
The final electron's entropy, $S^{EE}$, above has two contributions.  Its first two terms give the Shannon entropy for inclusive scattering while its third term is due to $S_k$.  $S_k$ is the scattered electron's momentum entropy assuming inclusive scattering occurs.  Notice that its last logarithm has dimensions of inverse momentum cubed.  $S_k$'s entropy arises from quantum mechanics ($\hbar$), position ($V=(2\pi\hbar)^3\delta^3(0)$), and momentum. The unregularized volume $V$ will not appear when calculating the expected momentum, expected helicity, and correlation or mutual information between the momentum and helicity (see \cite{shiva3}).\\  
\indent In \cite{pesch2}, eqn. (2.28), they calculate the momentum entanglement entropy for an elastically scattered particle, obtaining
\begin{align*}
\tilde{S}^{EE} = -\log(4\pi\hbar^2) +\log\sigma_T - \int dt\ \dfrac{1}{\sigma_{el}}\dfrac{d\sigma_{el}}{dt}\log\big(\dfrac{1}{\sigma_{el}}\dfrac{d\sigma_{el}}{dt} \big).
\end{align*}
Their $\tilde{S}^{EE}$ is akin to $S_k$ above in eqn. (\ref{sk}).  However, they have a $\log\sigma_T$ term.  Following our prescription in this subsection, their $\sigma_T$ would be $V/(\upsilon T)$, which they regularize as the total scattering cross section.  Since $\tilde{S}^{EE}$ assumes elastic scattering occured, just as $S_k$ above assumes inclusive scattering occurred, identifying their factor $V/(\upsilon T)$ with the elastic cross section, $\sigma_{el}$, would have been more appropriate.

%The volume, $V$, in eqn. (\ref{erho}) should be regularized.  Consider discretizing the sum ($\dfrac{V}{(2\pi\hbar)^3}\int d^3\vb*k\rightarrow \sum\limits_{\vb*k}$).  Then, the regularized volume is $V=\lim \limits_{\Delta\vb* k\rightarrow 0,\  n\rightarrow\infty}\sum_{m=1}^{n}\upsilon(\vb*k_0+n\Delta\vb*k)=.$

\section{Discussion}\label{discussion}

Unitarity constrains physical outcomes for general particle interactions. Suppose a pair of particles are entangled.  If one of them scatters from a third particle, the non-interacting particle's reduced density matrix $can$ change when not upholding unitarity (see $item$ ($\ref{v}$) in section \ref{unitarity}).  Another consequence of unitarity is it allows for easier calculations.  Subsection \ref{orbital} showed that unitarity and the language of density matrices allow for efficiently calculating the cross section for hard scattering without the scattering amplitude or Lippmann-Schwinger equation plus Green's function. Further consideration of unitarity may be warranted due to its importance as conveyed in this article.

The algorithms presented in this article can be readily applied to other interactions.  e.g., a full quantum information analysis of entanglement swapping between particles can be studied by first calculating the final density matrix of the particles' momenta and helicities.  With the density matrix in hand, finite correlations, expectation values, and mutual information of degrees of freedom such as helicities, momenta, and scattering angles can be computed.

%\section{Summary and conclusions}

\section*{Acknowledgements}
Thanks to Ms. Caroline Kibbe and Mr. Francis Simpatico of Brown University for their initial work on this project.  This research did not receive any specific grant from funding agencies in the public, commercial, or not-for-profit sectors.


%\begin{comment}
\appendix

\section{} \label{transition-deriv}

For hard (elastic) scattering of a spinless particle of mass $M$ and energy $E=\dfrac{(\hbar k)^2}{2M}$, prove the following relation between the transition matrix element and scattering amplitude up to a phase. \begin{align}\label{transition}
\la \vb*{k'}|\mathcal{T}|\vb*{k}\ra=\dfrac{\hbar^2}{(2\pi)^2M}f(\vb*{k'},\vb*{k})\ \delta(E_{\vb*{k'}}-E_{\vb*{k}})
\end{align}
Let the differential cross section in terms of the scattering amplitude be defined as $\dfrac{d\sigma}{d\Omega}\equiv|f(\vb*{k'},\vb*{k})|^{2}$.

$proof$:

Start with eqn. (\ref{apprho}), trace over the initial particle, and use the regularization eqn. (\ref{comptonreg}).  Denote the transition matrix $\la \vb*k'|\mathcal{T}|\vb*k\ra$ by $\mathcal{T}_{\vb*k,\vb*k'}$.  The final density matrix after hard scattering becomes
\begin{align*}
\rho^f &= \dfrac{1}{\delta^3(0)}\int d^3\vb*k' \int d^3\vb*k''\ (2\pi)^2 \mathcal{T}_{\vb*k,\vb*k''}\ (\mathcal{T}_{\vb*k,\vb*k'})^\dagger\ |\vb*k''\ra \la \vb*k'|\\
&= \dfrac{1}{\sigma} \int d^3\vb*k' \int d^3\vb*k''\ \dfrac{(2\pi)^4}{ \upsilon \hbar\ \delta(0)} \mathcal{T}_{\vb*k,\vb*k''}\ (\mathcal{T}_{\vb*k,\vb*k'})^\dagger\ |\vb*k''\ra \la \vb*k'|.
\end{align*}

The regularization or scattering cross section in the denominator above is $\sigma = \dfrac{V}{\upsilon T} = \dfrac{(2\pi)^3\delta^3(0)}{\upsilon\ (2\pi\hbar\ \delta(0))}$.  After tracing $\rho^f$,
\begin{align*}
1 &= \dfrac{1}{\sigma} \int d^3\vb*k' \dfrac{(2\pi)^4}{ \upsilon \hbar\ \delta(0)} |\mathcal{T}_{\vb*k,\vb*k'}|^2\\
& = \dfrac{1}{\sigma} \int d\Omega\ \ \Big(\int dk'\ |\vb*k'|^2 \dfrac{(2\pi)^4}{ \upsilon \hbar\ \delta(0)} |\mathcal{T}_{\vb*k,\vb*k'}|^2  \Big).
\end{align*}
The last equality implies that the factor in parentheses is the differential cross section. 
\begin{align}\label{dsigma}
\dfrac{d\sigma}{d\Omega} &= \int dk'\ |\vb*k'|^2 \dfrac{(2\pi)^4}{ \upsilon \hbar\ \delta(0)}\ |\mathcal{T}_{\vb*k,\vb*k'}|^2 \\
&\equiv |f(\vb*k',\vb*k)|^2\nn
\end{align}
Since the interaction is elastic, the transition amplitude, $\mathcal{T}_{\vb*k,\vb*k'}$, in eqn. (\ref{dsigma}) only contributes to the integral when $|\vb*k'| = |\vb*k|$.  This implies the relation 
\begin{align}\label{proposal}
\mathcal{T}_{\vb*k,\vb*k'} \equiv g(\vb*k')\ \delta(E_{\vb*k'} - E_{\vb*k}) 
\end{align}
for an unknown function $g(\vb*k')$.  

Plugging eqn. (\ref{proposal}) into eqn. (\ref{dsigma}) implies $g(\vb*k')$ equals $\dfrac{\hbar^2}{(2\pi)^2M}f(\vb*{k'},\vb*{k})  $ up to a phase and confirms eqn. (\ref{transition}).  Eqn. (\ref{transition})  would still occur if the Dirac delta function had been written as $\delta(k'-k)$ instead of $\delta(E_{\vb*k'} - E_{\vb*k})$ in eqn. (\ref{proposal}).  Without unitarity or the optical theorem and the language of density matrices suggesting the regularization, the above proof is not possible.

\bibliographystyle{elsarticle-harv} 

\bibliography{main}

%\begin{thebibliography}{00}

%% For example:

%\bibitem{atlas}
%G.~Aad \textit{et al.} [ATLAS],
%``Observation of quantum entanglement with top quarks at the ATLAS detector,''
%Nature \textbf{633}, no.8030, 542-547 (2024)
%doi:10.1038/s41586-024-07824-z
%[arXiv:2311.07288 [hep-ex]].
%91 citations counted in INSPIRE as of 01 Feb 2025

%\end{thebibliography}

\end{document}